\documentclass[epj,nopacs]{svjour}
\usepackage{graphics}
\usepackage{hyperref}
\usepackage{graphicx}
\usepackage{amsmath}
\usepackage{amssymb}
\usepackage{color}
\usepackage{comment}
\begin{document}
\title{Hydrodynamic instabilities in active cholesteric liquid crystals}
\author{Carl A. Whitfield\inst{1} \and Tapan Chandra Adhyapak\inst{2} \and Adriano Tiribocchi\inst{3} \and Gareth P. Alexander\inst{1,4} \and Davide Marenduzzo\inst{5} \and Sriram Ramaswamy\inst{6,7}\thanks{Present and permanent address.}
}                     
\institute{
Department of Physics, University of Warwick, Coventry CV4 7AL, UK \and 
Institut f\"{u}r Physik, Johannes Gutenberg-Universit\"at Mainz, Staudingerweg 7-9, 55128 Mainz, Germany \and
Dipartimento di Fisica e Astronomia, Universit\`a di Padova, Via Marzolo 8, I-35131 Padova, Italy \and
Centre for Complexity Science, University of Warwick, Coventry CV4 7AL, UK \and 
SUPA, School of Physics and Astronomy, University of Edinburgh, JCMB Kings Buildings, Mayfield Road, Edinburgh EH9 3JZ, Scotland \and 
TIFR Centre for Interdisciplinary Sciences, Tata Institute of Fundamental Research, 21 Brundavan Colony, Narsingi, Hyderabad 500 075 India \and
Department of Physics, Indian Institute of Science, Bangalore 560 012 India
}
\date{Received: date / Revised version: date}
%
\abstract{
We describe the basic properties and consequences of introducing active stresses, with principal direction along the local director, in cholesteric liquid crystals. The helical ground state is found to be linearly unstable to extensile stresses, without threshold in the limit of infinite system size, whereas contractile stresses are hydrodynamically screened by the cholesteric elasticity to give a finite threshold. This is confirmed numerically and the non-linear consequences of instability, in both extensile and contractile cases, are studied. We also consider the stresses associated to defects in the cholesteric pitch ($\lambda$ lines) and show how the geometry near to the defect generates threshold-less flows reminiscent of those for defects in active nematics. At large extensile activity $\lambda$ lines are spontaneously created and can form steady state patterns sustained by constant active flows. 
%
} 
\maketitle

\section{Introduction and Phenomenology}

Active liquid crystals have come to represent an archetype for active matter, offering a framework for organising ideas about biological processes and biologically inspired materials. Starting with studies of polar flocks~\cite{Vicsek1995,Toner1995,Toner2005}, the field has grown to encompass bacterial swarms and growing colonies~\cite{Dombrowski2004,Cisneros2007,Volfson2008}, systems of self-propelled rods~\cite{Narayan2007,Kudrolli2008}, the cell cytoskeleton and suspensions of the biopolymers that constitute it~\cite{Kruse2005,Prost2015,Sanchez2012,Keber2014}, among many other biologically inspired systems. From the phenomenology of liquid crystals certain generic traits of active matter have been identified, such as spontaneous flow transitions and hydrodynamic instabilities, giant number fluctuations and the role of defects in active turbulence~\cite{Ramaswamy2010,Marchetti2013}. From these a picture of living matter, viewed as a material, is emerging. However, from the perspective of condensed matter physics all of these studies represent only a small part of the broad scope offered by active matter phases, corresponding largely to the two-dimensional nematic and polar-ordered~\cite{Giomi2008,Sokolov2007,Aranson2007,Brotto2015} liquid crystal phases. Recently, active smectic phases have also been studied~\cite{Adhyapak2013,Chen2013,Romanczuk2016}, and there has been a parallel development of active matter without orientational order~\cite{Cates2015}, but active phases with symmetries different from simple nematic liquid crystals, or polar fluids, remain largely unexplored. 

Chirality is ubiquitous in nature, from the helical structure of DNA~\cite{Watson1953}, to bacterial flagella and their rotary motors~\cite{Berg1973,Berg2003}, to selective reflection and structural colour in both plants and animals~\cite{Srinivasarao1999,Kinoshita2005}. Although there has been some consideration of active stresses that are chiral~\cite{Furthauer2012,Furthauer2013}, such as those arising from torque-dipoles~\cite{Furthauer2013b}, active materials that have a chiral structure have not yet been studied. There is ample motivation to do so: First, from the structural point of view, it is noteworthy that cholesteric textures are found in a wide range of biological systems. Cross-sectional cuts through many fibrous tissues, such as dinoflagellate chromosomes, the carapaces of insects and crustaceans, fish eggshells and compact bones, display a distinctive series of arced fibrils~\cite{Bouligand2008,Bouligand1968,Bouligand1969,Bouligand1972}. These are the hallmark of a three-dimensional helical stacking of straight filaments identical to the structure of cholesteric liquid crystals. More significantly, polymers extracted and purified from such materials, as well as numerous other biopolymers, including actin, cellulose, chitin, collagen, microtubules, nucleic acids, polypeptides and polysaccharides all show cholesteric phases in solution~\cite{Bouligand2008,Robinson1961,Robinson1966,Bouligand1978,Livolant1984,Livolant1986,Livolant1991,Strzelecka1988,Yevdokimov1988,Revol1992,Revol1993}, and the same cholesteric ordering is found in colloidal suspensions of rod-like viruses, such as the {\em fd} virus~\cite{Lapointe1973,Dogic2000,Dogic2006}. Further examples of cholesteric order in biological materials include the packing of DNA in phage capsids~\cite{Arsuaga2005,Marenduzzo2009}, the origin of iridescence of Scarabaeidae beetle exoskeletons~\cite{Robinson1966,Michelson1911,Sharma2009}, the natural colour of flower petals~\cite{Zsila2001} and {\sl Pollia} fruit~\cite{Vignolini2012}, and silk spinning processes~\cite{Willcox1996}.  

Given these examples, it is natural to expect that active stresses in cholesteric phases should have broad relevance to a diverse range of biological and biologically inspired materials. Our motivations are also sparked by possibilities of experimental realisation of active cholesterics. All the natural systems that we have already mentioned do have such prospects: for instance, F-actin solutions actually exist in a cholesteric, rather than a nematic phase, as a consequence of the double helical, hence chiral, nature of an actin fibre. Another appealing experimental candidate for an active chiral gel is a solution of DNA molecules interacting with DNA or RNA polymerases, which lead to relative DNA-enzyme motion and may exert non-thermal active stresses on the polymers. That macroscopic chirality is important here is clear from looking at the passive counterpart of this system, namely a concentrated DNA solution without polymerases: such solutions have long been known to exhibit cholesteric or blue phases in different salt conditions or concentrations~\cite{Livolant1986,Livolant1991,Leforstier1994}. It is also worth noting that many passive cholesterics are formed by adding small amounts of a chiral dopant to nematic materials. This suggests that current experimental realisations of active nematics~\cite{Sanchez2012,Keber2014} could be converted to cholesterics by addition of suitable dopant, for instance a biopolymer known to form cholesteric mesophases, or {\em fd} virus. 

Here, we study active cholesterics from the framework of active liquid crystals, combining the force-dipole stresses already established for active materials with the cholesteric ordering that comes from a passive chiral nematic. One can expect chirality to lead to important effects in active materials, similar to the situation in passive cholesteric liquid crystals, whose hydrodynamics and physics are much different from those of nematics~\cite{deGennesProst}. With respect to the few existing theoretical works on chiral activity~\cite{Furthauer2012,Furthauer2013} our paper provides a systematic analysis of the active instabilities of the cholesteric ground state, as well as the active stresses and flows generated by defects in the cholesteric order, known as $\lambda$ lines. We do so by means of linear stability analysis and direct numerical solution of the non-linear equations of motion. Furthermore, we provide a detailed comparison between active cholesterics and the behaviour already established for active nematic and smectic phases. 

\begin{figure}
\centering
\includegraphics[width=\columnwidth]{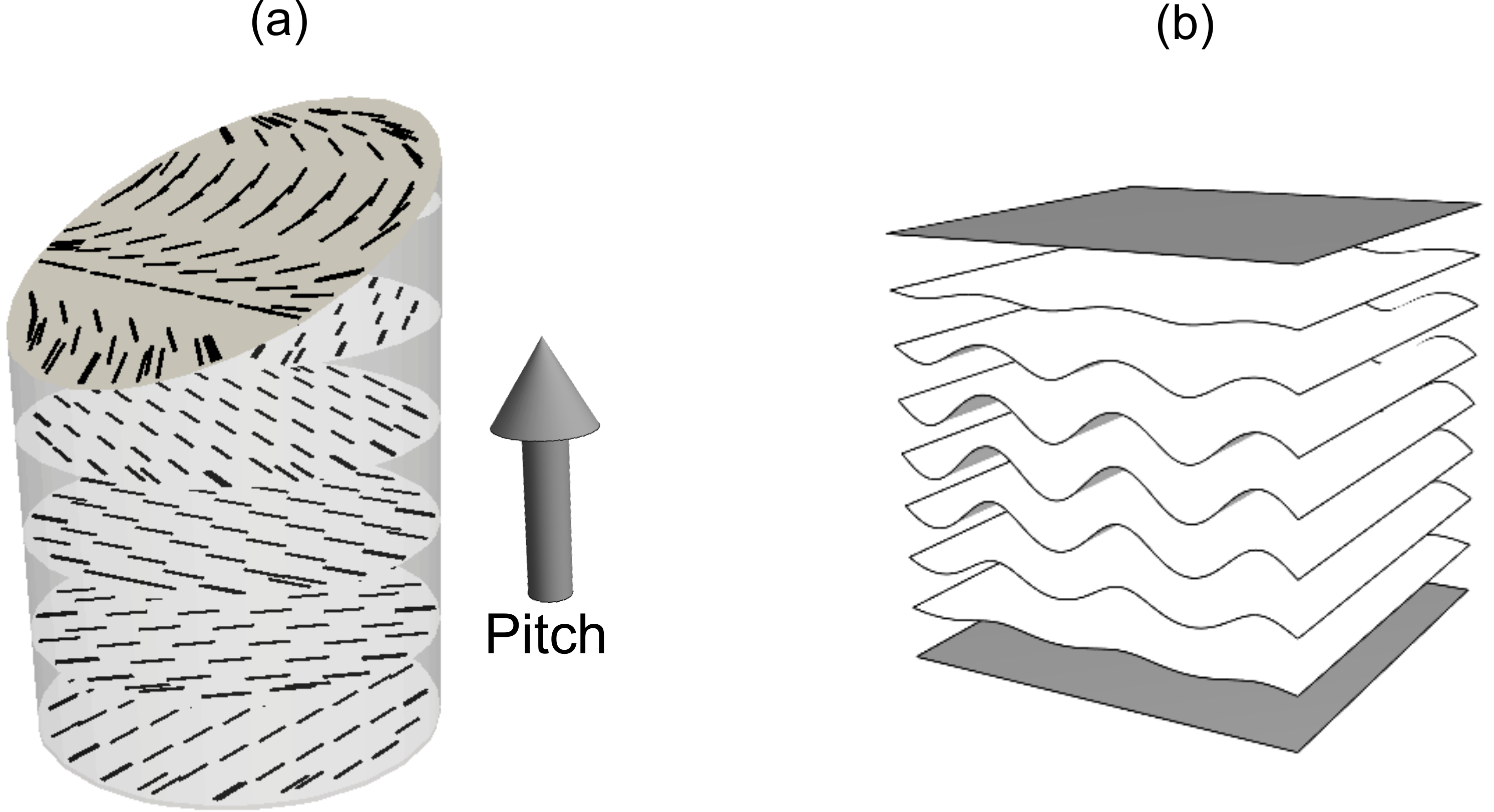}
\caption{(a) Projections of a cholesteric director field, equation \eqref{n0}, onto successive horizontal slices. The right-handed twist in the $z$ direction results in a texture of periodic arcs on an arbitrary slice (top slice). (b) Sketch of the Helfrich-Hurault instability of the cholesteric `layers' for a passive cholesteric under mechanical strain applied at the walls in the vertical direction.}
\label{fig:sketch}
\end{figure}

We first give a qualitative, phenomenological summary of active cholesterics and list our main results: Like nematics, cholesterics are described by a unit magnitude director field, $\hat{\vec{n}}$, corresponding to the local molecular alignment. Unlike nematics this is not uniform in the ground state, but is a linear function of position, the director rotating at a uniform rate, $q_0$, about a spontaneously chosen direction called the pitch axis. Denoting by $z$ this direction, the cholesteric ground state corresponds to the director 
\begin{equation}
\label{n0} \hat{\vec{n}} = \cos q_0z\, \hat{\vec{e}}_x + \sin q_0z\, \hat{\vec{e}}_y \equiv \hat{\vec{n}}_{0} ,
\end{equation}
and is shown in schematic form in figure~\ref{fig:sketch}. In terms of their fundamental description, nematics and cholesterics differ only in the presence of the chiral coupling constant, $q_0$, in the Frank free energy~\cite{deGennesProst,ChaikinLubensky}. At length scales that are short compared to the pitch, $\pi/q_0$, the two materials are effectively alike. Nonetheless, the bulk properties and characteristics of cholesterics differ substantially from those of nematics. At wavenumbers $k\ll q_0$ the structure has a one-dimensional periodicity analogous to the density modulation of smectics and hence its elasticity is the same~\cite{Toner1981}. However, in significant ways cholesterics and smectics are also fundamentally unalike. In the deepest sense this is because -- even if we ignore fluctuations -- the cholesteric state does not break translation invariance: the pitch axis is a continuous screw axis, an arbitrary translation along which can be compensated by a rotation about it. In the cholesteric the one-dimensional periodic modulation carries the director through a continuous family of symmetry-equivalent directions; no such symmetry relates the different densities encountered in the smectic mass-density wave. Whereas the layer normal in a smectic is parallel to the molecular alignment, the pitch axis is orthogonal to the director in a cholesteric. Geometry aids greatly in the description of liquid crystalline structures and their distortions. Several combinations of these with activity can be identified and it is useful to consider each separately. Active stresses in liquid crystals drive fluid flows in response to distortions in the director, with splay distortions generating flows locally parallel to the director and bend distortions creating perpendicular flows. 

The main results of our work are: (i) It is well established that contractile stresses drive an instability of the nematic state to splay deformations and extensile stresses yield instability to bend~\cite{Ramaswamy2010}. In passive liquid crystals, Lubensky has shown that splay distortions in the director are not true hydrodynamic modes in a cholesteric but are screened at length scales of order the cholesteric pitch~\cite{Lubensky1972}. Thus one may expect that the contractile instability of active nematics will be suppressed in a cholesteric until the activity exceeds a threshold of order $K q_0^2$ and indeed this is what we find. The Stokesian hydrodynamic deformations of a cholesteric involve bend distortions of the director and may be characterised either as a splay of the pitch direction, or a bend of the cholesteric `layers'. As these are true hydrodynamic modes, extensile active materials are still linearly unstable to them. This `layer undulation' instability is analogous in character to the Helfrich-Hurault effect (figure \ref{fig:sketch}(b)) observed in passive cholesterics and smectics in response to an applied electric field~\cite{Helfrich1970,Hurault1973}, or an extensional strain~\cite{Clark1973}. The same layer undulation instability also occurs in active smectics~\cite{Adhyapak2013}, but for contractile rather than extensile materials. The difference derives from the contrasting alignment of the director relative to the one-dimensional periodicity and emphasises that the active properties of cholesterics and smectics differ significantly even though their passive elasticity is the same. 

(ii) The direction associated to bend distortions is always orthogonal to the local director field. In a nematic all such orthogonal directions are equivalent, but this is not so in a cholesteric: one of the orthogonal directions corresponds to the cholesteric pitch axis, the direction along which the alignment is rotating. In the hydrodynamic mode associated to extensile instability, the bend distortion is directed along the pitch axis and generates flows in this direction. It is known that such flows are plug-like, rather than Poiseuille-like, in a passive cholesteric, and are resisted by an effective viscosity that can exceed that in nematics by five or six orders of magnitude~\cite{Helfrich1969,Lubensky1972,Marenduzzo2004}. 

(iii) In cholesterics, the case in which bend distortions are perpendicular to the pitch axis deserves separate study, both for geometric reasons and because the active flows that are generated lie within the cholesteric layers, where the effective viscosity is much lower. Such distortions and flows arise most simply in response to a uniform conical tilt of the cholesteric director into the pitch direction. This situation can be described in a quasi-one-dimensional setting identical to that studied extensively in the context of spontaneous flow transitions in active nematics~\cite{Marenduzzo2007} and polar gels~\cite{Voituriez2005}. In this sense it may be considered the natural cholesteric analogue of those studies and the analysis is broadly the same, although the structure of the director distortions and flows are markedly different, again emphasising the contrasting character of active cholesterics compared to other forms of active matter. The one-dimensional setting allows for exact solutions, beyond just linear analysis, although the restriction to an assumed one-dimensional variation also artificially suppresses the basic layer undulation instability so that such analysis gives only partial, qualitative insight. However, this is no different to the situation in active nematics, where such quasi-one-dimensional studies are still highly instructive~\cite{Marenduzzo2007,Voituriez2005,Marchetti2013}. 

(iv) Defects play a fundamental role in the non-linear description of materials, both passive and active. In active nematics, active stresses nucleate defects and drive their self-propulsion, eventually creating a turbulent dynamic steady state~\cite{Sanchez2012,Thampi2014a,Thampi2014b,Giomi2015}. The fundamental defects in cholesteric order are called $\lambda$ lines; they can be thought of as defects in the pitch, or dislocations in the cholesteric layers. The layer undulation instability ultimately gives way to the formation of pairs of $\lambda^{\pm}$ lines, reminiscent to the defect formation process in active nematics. A local analysis of the director field around a $\lambda$ line shows that they do not self-propel, as defects in active nematics do, but they still generate active flows in the cholesteric. We find that extensile active cholesterics can form steady state defect patterns in both bulk and confined geometries. The resulting structure is a lattice of $\lambda$ lines mediated by regions of cholesteric order, with a defect density that increases with activity. Ultimately, this ordered state is destroyed at larger activity with a transition to a state of active turbulence.

\section{Hydrodynamics of Active Liquid Crystals}
\label{sec:hydrodynamics}

The equilibrium elasticity of cholesteric liquid crystals is given by the Frank free energy for the director field 
\begin{equation}
\begin{split}
\label{Frank} F & = \int d^3r\, \biggl\{ \frac{K_1}{2} \bigl( \nabla \cdot \hat{\vec{n}} \bigr)^2 + \frac{K_2}{2} \bigl( \hat{\vec{n}} \cdot \nabla \times \hat{\vec{n}} + q_0 \bigr)^2 \\
& \hspace{18mm} + \frac{K_3}{2} \bigl( ( \hat{\vec{n}} \cdot \nabla ) \hat{\vec{n}} \bigr)^2 \biggr\} ,
\end{split}
\end{equation}
where $q_0$ is a chiral coupling constant that vanishes in a nematic. The free energy is minimised by a director with constant right-handed twist, as in \eqref{n0}. 

In addition to the director field, the hydrodynamic variables in an active cholesteric are the fluid mass density, $\rho$, and momentum density, $\vec{g}=\rho\vec{v}$, as well as the concentration, $c$, of active particles~\cite{Ramaswamy2010,Simha2002}. Their equations of motion are constructed by retaining, at leading orders in gradients, all terms allowed by symmetries and conservation laws and not necessarily derivable from a free energy. We will assume that the mass density and concentration of active particles are both homogeneous and constant, which implies that the fluid velocity is incompressible, $\nabla \cdot \vec{v} = 0$. In the viscous regime appropriate to represent active matter the continuity of the momentum density reduces to the Stokes equation, $\nabla \cdot \vec{\sigma} = 0$, where the stress tensor has components 
\begin{equation}
\begin{split}
\label{stress} \sigma_{ij} & = - P \delta_{ij} + 2 \eta u_{ij} + \frac{\nu}{2} \bigl( n_i h_j + h_i n_j \bigr) \\ 
& \quad + \frac{1}{2} \bigl( n_i h_j - h_i n_j \bigr) - \Phi_{ik}\partial_j n_k - \zeta n_in_j . 
\end{split} 
\end{equation}
Here, $P$ is the pressure, $\eta$ the viscosity and $u_{ij}=(\partial_i v_j + \partial_j v_i)/2$ the symmetric part of the velocity gradients of the fluid. $\vec{h}=-\delta F/\delta \hat{\vec{n}}$ is the molecular field conjugate to the liquid crystal director, and $\nu$ is a flow alignment parameter. The liquid crystal is flow aligning if $|\nu|\geq 1$ and flow tumbling if $|\nu|<1$~\cite{deGennesProst}. We choose $\nu<-1$ corresponding to a flow-aligning active cholesteric with rod-like particles ({\sl e.g.} F-actin filaments or DNA molecules). The ``Ericksen stress'' term contains $\Phi_{ij} = [\partial f / \partial (\partial_i n_k)](\delta_{jk} - n_j n_k)$ where $f$ is the free energy density (such that $F=\int{\rm d}^3r f$ in equation \eqref{Frank}). The final term is the active stress, which corresponds to a force dipole aligned along the local director field of the liquid crystal~\cite{Simha2002}. The phenomenological coefficient $\zeta$ is proportional to the concentration of active particles and is positive in extensile materials and negative in contractile ones. The active stress ultimately arises through an off-diagonal piece in the matrix of kinetic coefficients linking stress and fuel consumption, viewed as `fluxes', to strain-rate and chemical-potential imbalance, viewed as `forces'~\cite{Marchetti2013}. 

Finally, the dynamic equation for the relaxation of the director field is 
\begin{equation}
\label{dndt} \partial_t n_i + v_j \partial_j n_i + \omega_{ij} n_j = - \nu u_{ij} n_j + \frac{1}{\gamma} h_i ,
\end{equation}
where $\omega_{ij}=(\partial_i v_j - \partial_j v_i)/2$ is the antisymmetric, or rotational, part of the velocity gradients, and is the same as in a passive liquid crystal. The parameter $\gamma$ is a rotational viscosity which sets the timescale for reorientation due to the molecular field. 

\section{Linear Instabilities in an Active Cholesteric}

\subsection{Generic hydrodynamic instability: pitch-splay} 
\label{sec:hydroinst}

The hydrodynamics of cholesterics is subtle, as not all director deformations correspond to true hydrodynamic modes. We follow the framework of Lubensky~\cite{Lubensky1972} for the hydrodynamics of passive liquid crystals to calculate hydrodynamic instabilities in active cholesterics in the Stokesian regime. A generic perturbation of the cholesteric ground state~\eqref{n0} can be written  
\begin{align}
\begin{split}
\hat{\vec{n}} & = \cos\delta\theta \bigl[ \cos (q_0z+\delta \phi) \hat{\vec{e}}_x + \sin (q_0z+\delta \phi) \hat{\vec{e}}_y \bigr] \\
& \quad + \sin\delta\theta \hat{\vec{e}}_z , 
\end{split} \\ 
\label{ndir} & \approx \hat{\vec{n}}_0 + \delta \phi \, \hat{\vec{n}}_{\perp 0} + \delta\theta\, \hat{\vec{e}}_z ,
\end{align} 
where $\hat{\vec{n}}_{\perp 0} = -\sin q_0 z\,\hat{\vec{e}}_x + \cos q_0 z\,\hat{\vec{e}}_y$, and $\delta \phi = \delta \phi(x,z,t)$ and $\delta\theta = \delta \theta(x,z,t)$ are small fluctuations. 
Without loss of generality we consider the mode with wavevector $(k_x,0,k_z)$ lying in the $x$-$z$ plane, which, due to the symmetry of the cholesteric ground state, can be written as a sum over all Brillouin zones 
\begin{align}
\label{dp} \delta \phi(x,z,t) = \sum_n \delta \phi_n e^{i[k_x x + (k_z + nq_0)z - \omega t]} \\
\label{dq} \delta \theta(x,z,t) = \sum_n \delta \theta_n e^{i[k_x x + (k_z + nq_0)z - \omega t]}
\end{align}
where $k_z\in[-\pi/q_0,\pi/q_0]$ is in the first Brillouin zone. Henceforth all terms with subscript $n$ are understood to denote the $n$th component of the infinite sum over all Brillouin zones. 

The molecular field $\vec{h}_n$ (given in Appendix \ref{app:flow}) is coupled to the modes $\delta \phi_{n\pm2}$ and $\delta \theta_{n\pm1}$. Indeed, the symmetry of the cholesteric basis means that even $n$ modes of $\delta \phi_n$ are coupled to odd $n$ modes of $\delta \theta_n$ and \emph{vice versa} \cite{Lubensky1972}. It is convenient to solve the Stokes equation in the Cartesian basis. The active contribution to the Stokes equation is
\begin{align}
\begin{split}
\label{fa} &\vec{f}^a_n \approx -\frac{\zeta}{2} \biggl\{ ik_x(\delta\theta_{n+1} + \delta \theta_{n-1}) \hat{\vec{e}}_z\\
& + \Bigl[ k_x(\delta \phi_{n+2} - \delta \phi_{n-2}) + i(k_z + n q_0)(\delta \theta_{n+1} + \delta \theta_{n-1}) \Bigr]\hat{\vec{e}}_x\\
& + \Bigl[ ik_x(\delta \phi_{n+2} + \delta \phi_{n-2}) - (k_z + n q_0)(\delta \theta_{n+1} - \delta \theta_{n-1}) \Bigr]\hat{\vec{e}}_y \biggr\} \,,
\end{split}
\end{align}
and interestingly does not depend on $\delta \phi_n$ but rather the modes in neighbouring Brillouin zones. Similarly, the passive distortion contribution to the Stokes equation $\vec{f}_n^d$ can be written in terms of the molecular field in equations, and is given in Appendix \ref{app:flow}. The stability of the initial perturbations $\delta \phi_n$ and $\delta \theta_n$ is determined by equation \eqref{dndt} and is given by 
\begin{align}
\begin{split}
\label{dpdt} -i\omega \delta \phi_n & = -q_0 v_n^{(z)} + \frac{ik_x}{4}\biggl[ i\nu \Bigl( v_{n+2}^{(x)} - v_{n-2}^{(x)} \Bigr)\\
& \quad + 2 v^{(y)}_n - \nu \Bigl( v_{n+2}^{(y)} + v_{n-2}^{(y)} \Bigr) \biggr] + \frac{h_n^{\perp 0}}{\gamma}
\end{split} \\
\begin{split}
\label{dqdt} -i\omega \delta \theta_n & = -\frac{\nu+1}{4}\biggl\{ \Bigl[ k_z + (n+1)q_0 \Bigr] (iv_{n+1}^{(x)} - v_{n+1}^{(y)}) \\
& \quad + \Bigl[ k_z + (n-1)q_0 \Bigr] (iv_{n-1}^{(x)} + v_{n-1}^{(y)})\biggr\}\\
& \quad - \frac{\nu - 1}{4}ik_x\Bigl(v_{n+1}^{(z)} + v_{n-1}^{(z)}\Bigr) + \frac{h_n^{z}}{\gamma} \, .
\end{split}
\end{align}
In general $\omega$ is the solution to an infinite-dimensional eigenvalue problem \cite{Lubensky1972} where each mode $n$ is coupled to neighbouring modes up to $n \pm 6$. In the hydrodynamic limit $k \ll q_0$ the lowest energy modes decouple at order $k^0$ to a $5 \times 5$ eigenvalue problem involving the modes $\delta \phi_0, \delta \theta_{\pm 1}$, and $\delta \phi_{\pm 2}$, given explicitly in Appendix~\ref{app:flow}. We look first for hydrodynamic modes of the passive case which correspond to eigenvalue solutions of $\omega \rightarrow 0$ as $\zeta, k \rightarrow 0$. We find that these exist in two cases, when $k_x = 0$ or $k_z = 0$. 

When $k_x = 0$ the matrix is diagonalised and the system becomes quasi-one-dimensional. In this case the only hydrodynamic mode has eigenfunction $\delta \phi_0$ and is purely diffusive with $\omega =  - i K_2 k_z^2/\gamma$. This mode remains unaffected by non-zero activity and so is not generically unstable in an active cholesteric.

\begin{figure}
\centering
\includegraphics[width=\columnwidth]{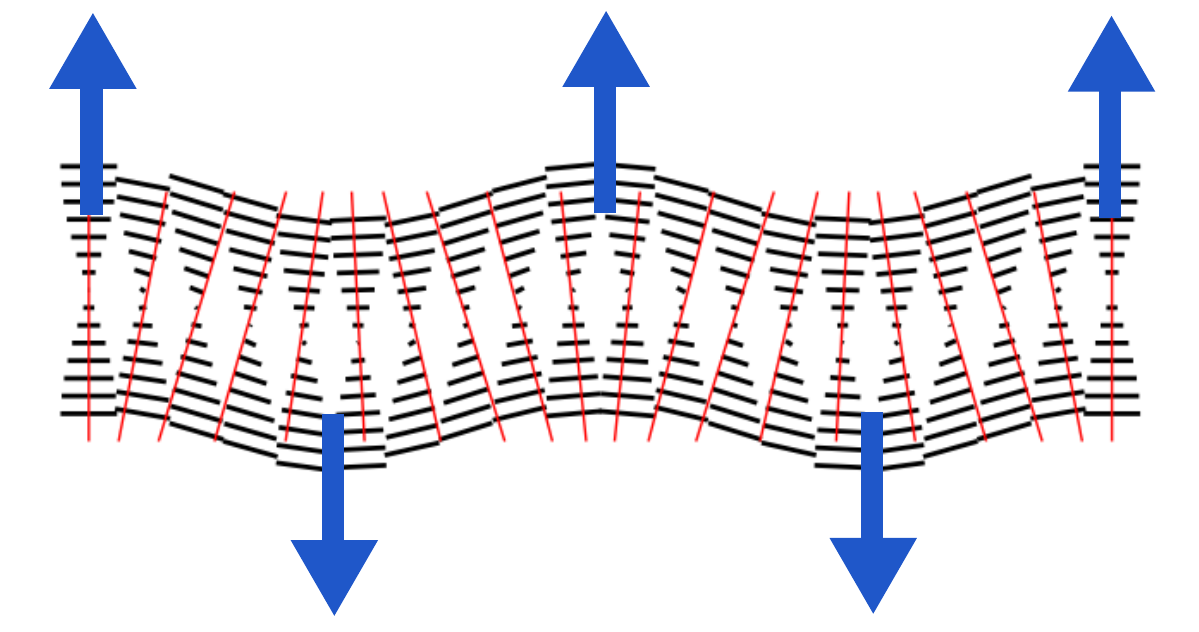}
\caption{Sketch of the pitch-splay, or layer undulation, instability in extensile cholesterics. Black lines show the projection of the twisted director field onto the plane containing the splayed pitch axis (red lines). The blue arrows show the active flow direction which acts to increase the distortion and drives the instability.}
\label{fig:modes}
\end{figure}

When $k_z = 0$ the problem can be further reduced to a system of three coupled equations for $\delta \phi_0$ and $\delta \theta_{\pm 1}$ in the long wavelength limit $k_x \ll q_0$. In the passive case ($\zeta = 0$) we find that the lowest energy mode is diffusive and has corresponding director perturbation:
\begin{align}
\label{eigf1} \delta \phi &= \delta \phi_0 e^{i (k_x x - \omega t)} \\
\label{eigf2} \delta \theta &= -\frac{i k_x}{q_0} \delta \phi_0 \cos(q_0 z) e^{i (k_x x - \omega t)}
\end{align}
which agrees with the analysis of the static equations by Lubensky \cite{Lubensky1972}. Including a small activity $|\zeta| \ll K q_0^2$ this mode becomes generically unstable to extensile activity with
\begin{align}
\label{omega0} \omega_0 \approx \frac{i}{\eta} \biggl[ \frac{\zeta}{2} - \frac{3K_3}{8}k^2 \biggr]\frac{8\eta + \gamma(1+\nu)^2}{8\eta + 2\gamma(1+\nu)^2} \, .
\end{align}
A sketch of this perturbation mode and the accompanying active flow is shown in figure~\ref{fig:modes}. As shown in \cite{Lubensky1972} this mode corresponds to a pure splay deformation of the pitch, or a bend of the director field that is directed along the pitch axis. When the activity is extensile, this deformation mode gives rise to active flows along the pitch axis, parallel to the bend direction, which destabilise the cholesteric order.
This hydrodynamic pitch-splay mode in a cholesteric is analogous to the layer undulation mode of smectic A materials. Indeed, their passive elasticity is the same~\cite{Toner1981}. However, the director distortions that underlie the cholesteric pitch-splay mode are of a different type than those associated to the layer undulations of a smectic. As we have described, in a cholesteric the director distortions are of bend type as this leads only to splay deformations of the pitch axis. In contrast, the director distortions of smectic layer undulations are splay deformations, since the director corresponds to the layer normal. It is for this reason that the instability of active cholesterics occurs for extensile materials, in contrast to the situation in active smectics, where it is the contractile material that is unstable~\cite{Adhyapak2013}. 

Modes which are destabilised by contractile activity ($\zeta < 0$) all have finite stability thresholds in the limit $k = 0$ of order $K q_0^2$. Thus one expects contractile cholesterics to be stable unless the magnitude of the activity exceeds (approximately) $K q_0^2$. This applies to $\omega_0$ which gives an instability for large enough contractile activity, but in general this may not by the lowest instability threshold mode. For example, the offset family of perturbations $\delta \theta_{2n}, \delta \phi_{2n-1}$ also predict finite instability thresholds in the long wavelength limit. In the following section we consider a special case of these in the quasi-one dimensional limit.

\subsection{Suppression of splay instability and spontaneous flow transition}
\label{sec:nemlimit}

Contractile active nematics are linearly unstable to splay distortions in the director field~\cite{Ramaswamy2010,Simha2002,Ramaswamy2007}. For a director lying in the $xy$-plane these splay distortions may be written in the form 
\begin{equation}
\label{nempert} \delta \vec{n} = \delta\theta\, \textrm{e}^{i(kz-\omega t)}\, \hat{\vec{e}}_z .
\end{equation} 
When $k \gg q_0$ the cholesteric ground state~\eqref{n0} locally resembles a nematic, on length scales of this perturbation, and the distortion retains its splay-like character. However, for $k\sim q_0$ the cholesteric order ``screens'' the instability and produces a finite threshold for its onset. 

Substituting \eqref{nempert} into the force balance equations we find that the fluid velocity is
\begin{equation}
\begin{split}
\vec{v} & = \delta \theta\, \textrm{e}^{i(kz-\omega t)} \biggl[ \zeta + \frac{1+\nu}{2} \bigl( K_1 k^2 + K_3 q_0^2 \bigr) \biggr] \\
& \qquad \times \frac{q_0 \hat{\vec{n}}_{\perp 0} - ik \hat{\vec{n}}_0}{\eta (k^2-q_0^2)} .
\end{split}
\end{equation}
Substituting this into \eqref{dndt} and keeping terms to first order in the perturbations we find the dispersion relation 
\begin{equation}
\label{nemstab} \omega = -i\zeta \frac{1+\nu}{2\eta} - i \bigl( K_1 k^2 + K_3 q_0^2 \bigr) \biggl[ \frac{1}{\gamma}+\frac{(1+\nu)^2}{4\eta} \biggr] .
\end{equation}
The active term is identical to that for a splayed perturbation in an active nematic~\cite{Ramaswamy2007}, and again is long range. The passive splay contribution proportional to $K_1$ also remains unchanged, and in the nematic limit $q_0\to 0$ we recover the familiar splay instability of active gels. The new contribution arising from the cholesteric order is the passive bend term proportional to $K_3 q_0^2$. We see that this is always negative and imaginary, and hence acts to screen the nematic splay mode. Instability only sets in when the activity exceeds a threshold of order $K_3q_0^2$. As in the nematic case~\cite{Voituriez2005,Marenduzzo2007}, for flow-aligning, rod-like particles ($\nu<-1$) the instability occurs for extensile activity ($\zeta>0$). 

\begin{figure}
\centering
\includegraphics[width=\columnwidth]{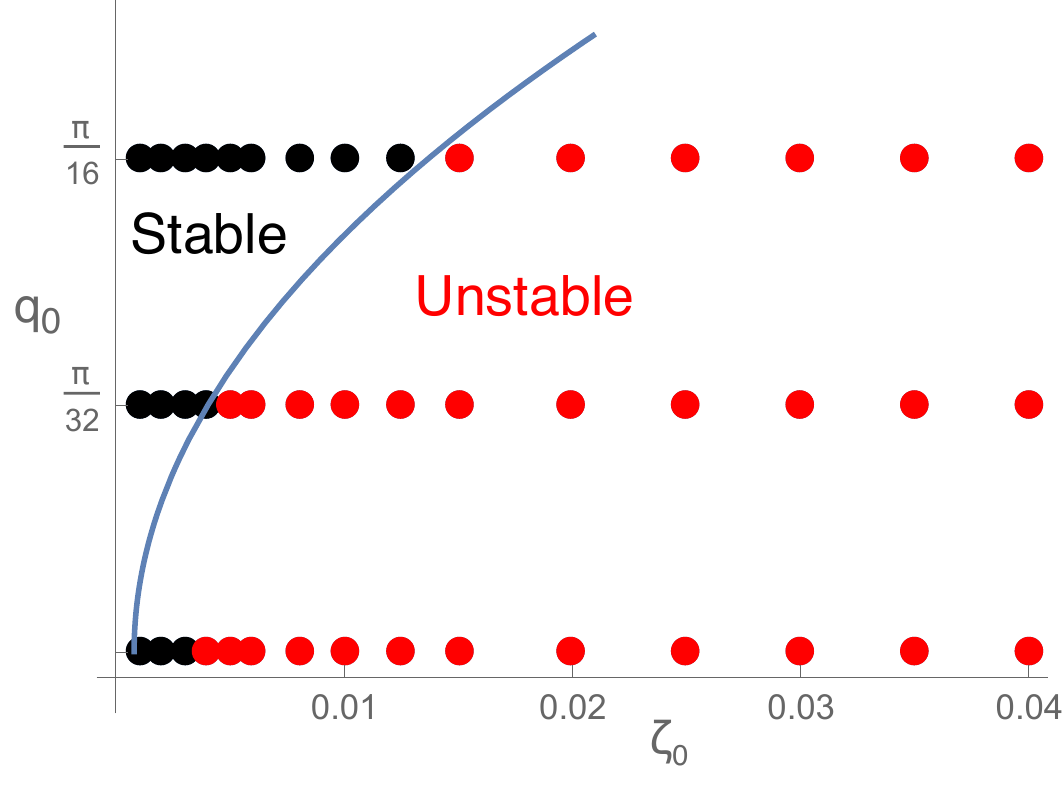}
\caption{Quasi-one-dimensional numerical results for extensile active cholesterics with three values of the pitch. The active gel is sandwiched between two infinite plates, parallel to the $xy$ plane, lying at $z=0$ and $z=L_z$. Planar boundary conditions are set on both walls. The line is the analytical prediction from \S~\ref{sec:nemlimit} for $k=\pi/L_z$, demonstrating the $q_0^2$ dependence of the activity threshold for this mode. The discrepancy at small values of $q_0$ is due to the fixed anchoring boundary conditions used in the simulations.}
\label{fig:quasi_1D}
\end{figure}

\begin{figure*}
\centering
\includegraphics[width=2\columnwidth]{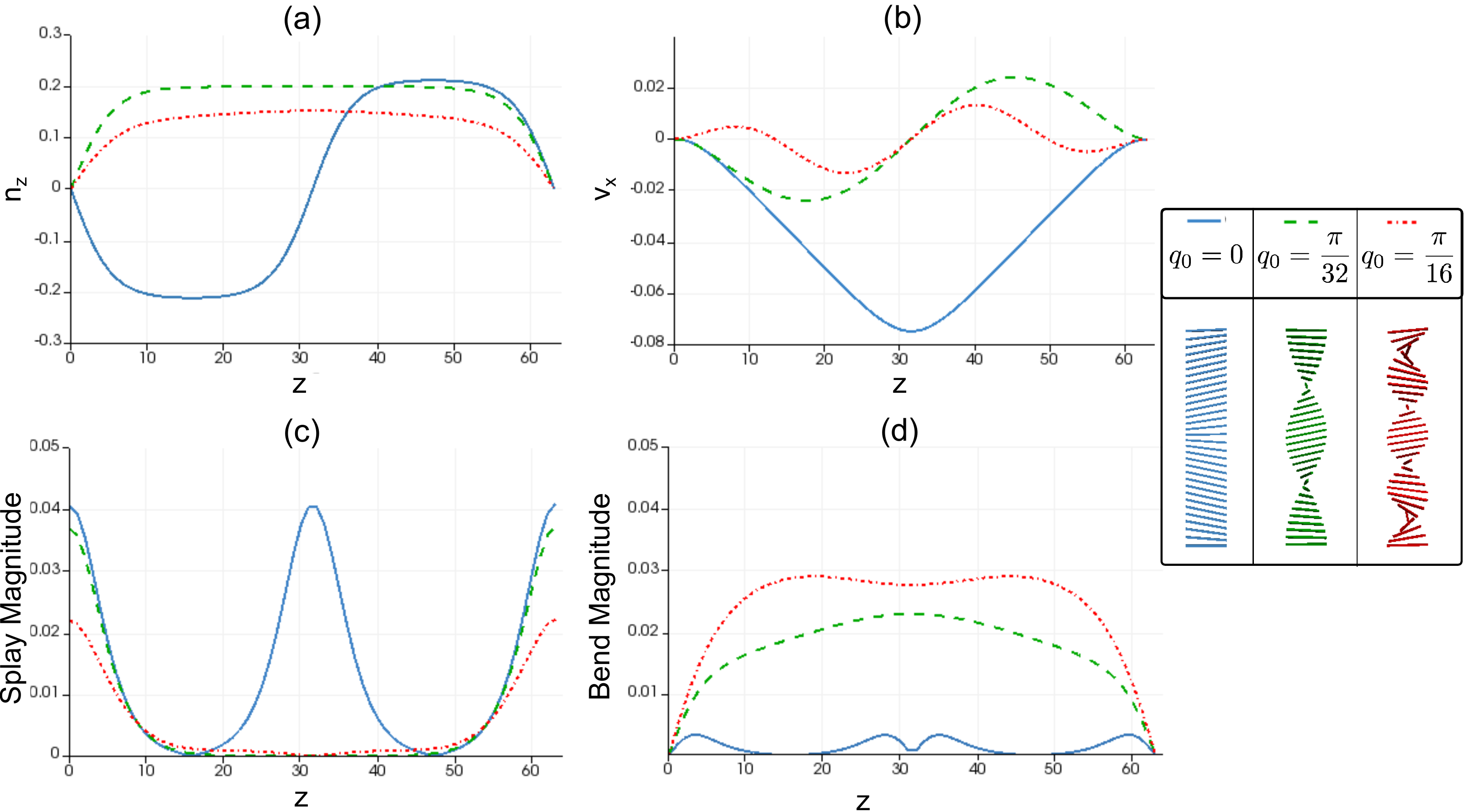}
\caption{Profiles of (a) director tilt, $n_z$, and (b) velocity along the $x$-direction, $v_x$, for three different steady states found for extensile active cholesterics ($\zeta=0.025$). Panels (c) and (d) show the magnitude of splay and bend deformations, respectively, for each of these steady states. $q_0=0$ is the nematic case, shown for reference. The legend on the right also shows schematics of the steady state director field in each case.} 
\label{fig:flow_dir_1D}
\end{figure*}

This linear analysis is confirmed by full numerical solution of the hydrodynamic equations for such a splay mode in an active cholesteric, as we show in figure~\ref{fig:quasi_1D}. As there is only $z$-dependence of both the perturbation and the cholesteric ground state, the simulations can be performed in a quasi-one-dimensional setting, although we defer details of the numerical method until \S~\ref{sec:simulations}. This set up is, therefore, entirely analogous to that considered in the context of spontaneous flow transitions in active nematics~\cite{Voituriez2005,Marenduzzo2007}. Also in the cholesteric case, there is a spontaneous flow transition above the threshold for linear instability, however its character is rather different. Figure~\ref{fig:flow_dir_1D} shows the results of numerical simulations of this transition in an extensile cholesteric confined between parallel plates with tangential anchoring conditions, as well as the analogous results for a nematic~\cite{Marenduzzo2007}, reproduced for comparison. In the nematic, the spontaneously flowing state is characterised by director splay distortions localised around the mid-plane of the cell, and also at the boundaries to accommodate the tangential anchoring. The bend distortions are uniformly small by comparison. The flow is predominantly along the director field, in conformity with the splay nature of the distortions driving them.  

By contrast, in the cholesteric the spontaneous flow transition leads to an approximately uniform conical tilt of the director along the pitch axis throughout the bulk of the cell, vanishing only near the boundaries to satisfy the tangential anchoring conditions. This leads to a heliconical director profile in which the bend distortion has constant magnitude and is directed orthogonal to the pitch axis. There is no splay, except close to the boundaries where it is due to the surface anchoring. The spontaneous flow is predominantly along the direction of the bend distortion and hence parallel to the vector $\hat{\vec{n}}_{\perp 0}$. Throughout the bulk of the cell it has approximately constant magnitude, but helically varying direction. Although we do not show it, simulations with a fully periodic cholesteric texture and no boundary conditions yield a spontaneous flow transition to a perfect heliconical director with a constant conical tilt angle. 

These simulations suggest a simple one-dimensional analysis of the flowing state analogous to that of Voituriez {\sl et al.}~\cite{Voituriez2005} for polar gels, but for the heliconical director field 
\begin{equation}
\label{heliconical} \hat{\vec{n}} = \cos\theta\, \hat{\vec{n}}_{0} + \sin\theta\, \hat{\vec{e}}_z ,
\end{equation}
with the conical angle $\theta$ a constant parameter to be solved for. There is no splay in this texture, but the bend vector is non-zero 
\begin{equation}
(\hat{\vec{n}} \cdot \nabla) \hat{\vec{n}} = q_0 \sin\theta \cos\theta\, \hat{\vec{n}}_{\perp 0} ,
\end{equation}
and of constant magnitude, oriented perpendicularly to both the director field and the pitch axis. The active stress associated to it drives a flow in the same direction 
\begin{equation}
\begin{split}
\vec{v} & = - q_0 \sin\theta \cos\theta \biggl( \frac{\zeta}{\eta q_0^2} + \frac{1 + \nu \cos 2\theta}{2\eta} \\
& \quad \qquad \times \bigl[ K_2 + (K_3-K_2) \cos 2\theta \bigr] \biggr) \hat{\vec{n}}_{\perp 0} .
\end{split}
\end{equation}
Finally, the angle $\theta$ is determined by the director relaxation equation~\eqref{dndt}. In addition to the passive solution $\theta=0$, the activity allows for a solution with non-zero conical tilt, given implicitly by 
\begin{equation}
\begin{split}
- \frac{\zeta}{2\eta q_0^2} & = \biggl( \frac{1}{\gamma \bigl( 1+\nu \cos 2\theta \bigr)} + \frac{\bigl( 1 + \nu \cos 2\theta \bigr)}{4\eta} \biggr) \\
& \qquad \times \bigl[ K_2 + (K_3-K_2) \cos 2\theta \bigr] .
\end{split}
\end{equation}

We conclude that in this simplified one-dimensional picture, the cholesteric is primarily unstable to bend deformations resulting from a constant tilt of the director along the pitch axis. Thus the splay instability of an active nematic is replaced by a bend instability in the active cholesteric. The bend vector twists with the director field and thus generates flows of constant magnitude along directions perpendicular to the pitch axis that rotate along it in a helical fashion. In view of its nature, we refer to this remarkable active flowing state as a ``sliding cholesteric''.

\section{Numerical Simulations}
\label{sec:simulations}

To study the nature of the state that develops from the cholesteric ground state in response to the fundamental hydrodynamic pitch-splay instability, we solve the non-linear equations of motion numerically. {More specifically, we use a hybrid lattice-Boltzmann (LB) algorithm, previously used for passive or active nematic liquid crystals~\cite{Cates2009,Tiribocchi2011,Tiribocchi2013}, to solve~\eqref{dndt}, and the following Navier-Stokes equation,
\begin{equation}\label{navierstokes}
\rho(\partial_t+ v_j \partial_j)v_i = \partial_j (\sigma_{ij})+\eta \partial_j(\partial_jv_i + \partial_i v_j),
\end{equation}
where all terms were defined previously (see Section~\S~\ref{sec:hydrodynamics}). The molecular field $h_i$ is defined in terms of the following free energy,
\begin{equation}
\begin{split}
\label{freeenergysim} F & = \int d^3r\,\biggl\{ \frac{\alpha}{2}{\vec{n}}^2+\frac{\beta}{4} \bigl({\vec n}^2\bigr)^2 + \frac{K_1}{2} \bigl( \nabla \cdot {\vec{n}} \bigr)^2 \\ 
 & + \frac{K_2}{2} \bigl( {\vec{n}} \cdot \nabla \times {\vec{n}} + q_0 \bigr)^2 
 + \frac{K_3}{2} \bigl( ( {\vec{n}} \cdot \nabla ) {\vec{n}} \bigr)^2 \biggr\}.
\end{split}
\end{equation}
We note that in our simulations $\vec{n}$ is not a unit vector; this constraint is though enforced softly through the bulk free energy which is minimised by $\vec{n}^2=-\alpha/\beta$ (we choose $\alpha=-\beta$, see below).}

We use our method to simulate an active cholesteric, either in a quasi-one-dimensional geometry (Figures~\ref{fig:quasi_1D} and \ref{fig:flow_dir_1D}), or in a quasi-two-dimensional geometry ($xz$-plane). The typical system size is $64$ (quasi-1D geometry) or $64\!\times\! 64$ (quasi-2D geometry) lattice sites with either periodic boundary conditions or homogeneous (planar) strong anchoring of the director field on both the upper and the lower wall bounding the simulation domain. The initial condition is the equilibrium helix~\eqref{n0}, for a suitable choice of $q_0$, with a small deformation in the midplane and no flow. In our calculations we have chosen $\alpha=-0.1$, $\beta=0.1$, $\gamma=1$, $\nu=-1.1$ (corresponding to a rod-like, flow-aligning liquid crystal), and $\eta=5/3$ as in previous numerical works~\cite{Tiribocchi2011,Tiribocchi2013}. We have also assumed the one elastic constant approximation, setting $K_1=K_2=K_3=K=0.01$. In the contractile case, for instance, these values can be mapped as done in~\cite{Tjhung2012} onto an actomyosin gel with effective elastic constant equal to 1 nN, $\gamma=1 \mathrm{kPa}\cdot \mathrm{s}$, and $\eta=1.67\mathrm{kPa}/\mathrm{s}$. When using walls, the value of the surface anchoring was chosen so as to be always in the strong anchoring limit.

\begin{figure*}
\centering
\includegraphics[width=1.6\columnwidth]{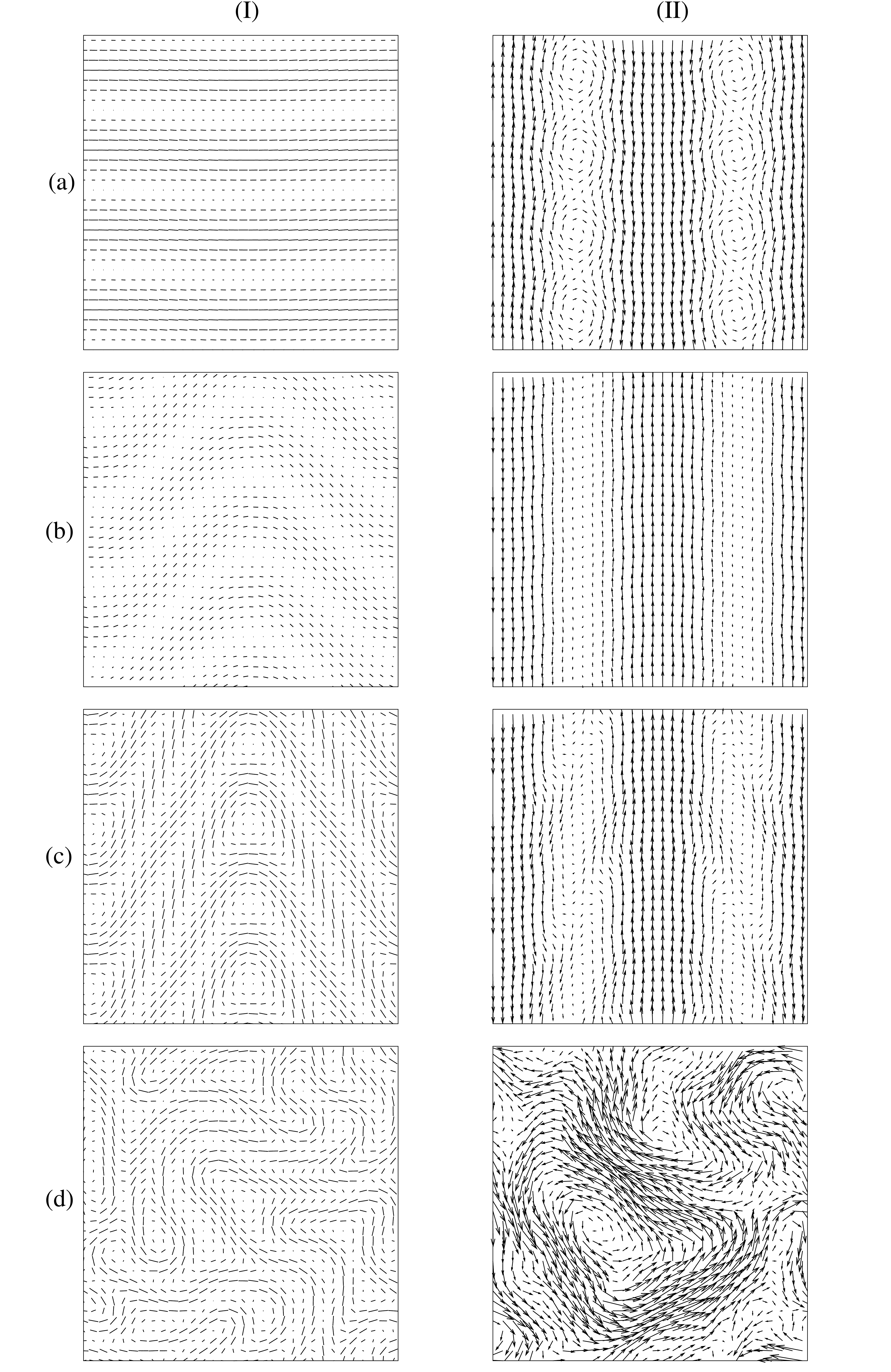}
\caption{Simulation results for an extensile active cholesteric in a quasi-two-dimensional geometry. The snapshots correspond to director field profile (I) and the corresponding velocity profile (II) for $\zeta$ equal to: $0.00009$ (a), $0.00025$ (b), $0.001$ (c), $0.005$ (d). In (a)-(c) these profiles are steady states of the system, whereas (d) is a representative snapshot of the time-dependent asymptotic state. All results are obtained for $q_0=\pi/16$.} 
\label{fig:flow_dir_2D_ext}
\end{figure*}

\begin{figure*}
\centering
\includegraphics[width=1.6\columnwidth]{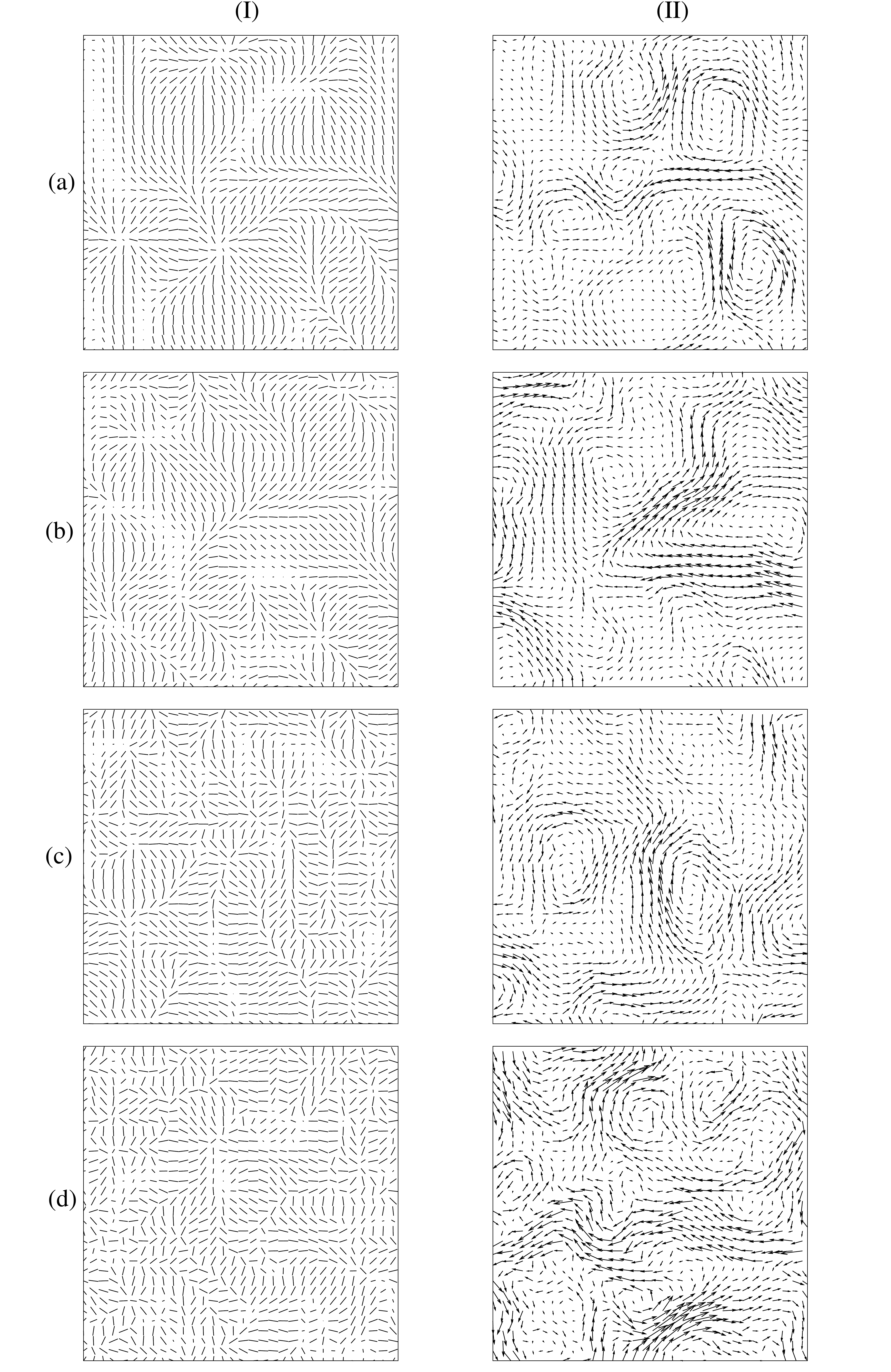}
\caption{Simulation results for a contractile active cholesteric in a quasi-two-dimensional geometry. The snapshots correspond to director field profile (I) and the corresponding velocity profile (II) for $\zeta$ equal to: $-0.01$ (a), $-0.015$ (b), $-0.03$ (c), $-0.05$ (d). In all (a)-(d) the snapshots are representative of a time-dependent asymptotic (statistically steady) state. All results are obtained for $q_0=\pi/16$.}
\label{fig:flow_dir_2D_contr}
\end{figure*}

\begin{figure*}
\centering
\includegraphics[width=1.6\columnwidth]{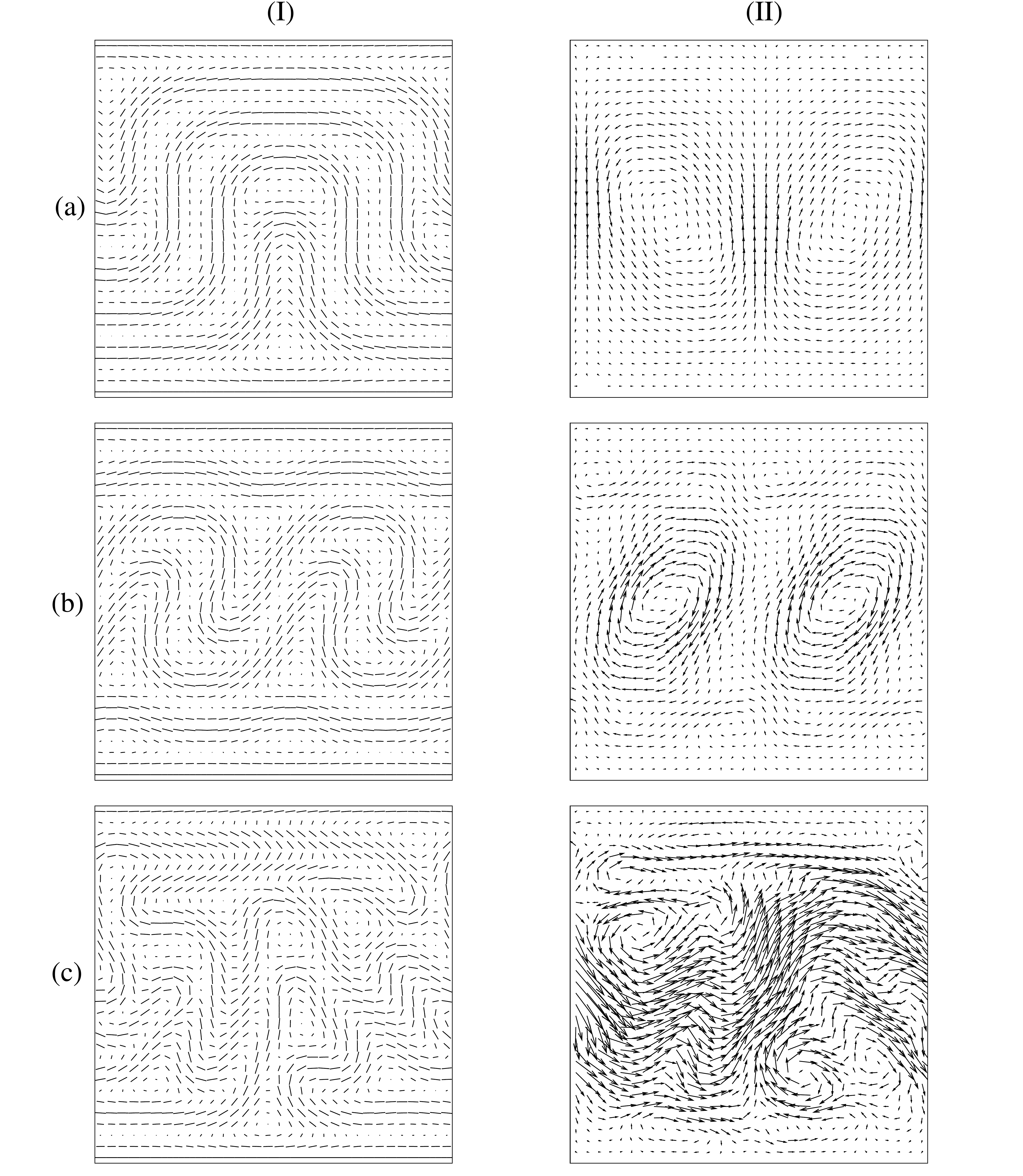}
\caption{Results of an extensile active cholesteric confined in a quasi-two-dimensional geometry with flat walls. Homogeneous anchoring of the director is set on both the upper and lower surfaces. The snapshots correspond to director field profile (I) and the corresponding velocity profile (II) for $\zeta$ equal to: $0.001$ (a), $0.0025$ (b), $0.005$ (c). In (a) and (b) these profiles are steady states of the system, whereas (c) is a snapshot of the time-dependent asymptotic state. All results are obtained for $q_0=\pi/16$.}
\label{fig:flow_dir_2D_ext_walls}
\end{figure*}

In the case of extensile activity, there is a continuous flow transition with the activity $\zeta$. Figure~\ref{fig:flow_dir_2D_ext} shows successive steady states for increasing extensile activities. As predicted in \S~\ref{sec:hydroinst} for an infinite cholesteric, we observe a pitch-splay instability as the director field bends and is advected along the direction of positive curvature as sketched in figure~\ref{fig:modes} for the linear instability. The bifurcation to spontaneous flow is supercritical, corresponding to a continuous nonequilibrium transition, as found in~\cite{Tjhung2011} for active nematics. The transition threshold is system size dependent and goes to zero in an infinite system (corresponding to $k\to 0$ in the linear stability condition~\eqref{omega0}). As the activity is increased the resulting bend deformations become more pronounced and vortices begin to appear in the flow (figures~\ref{fig:flow_dir_2D_ext}(a)-(c)). At larger activities still, $\lambda$ lines (defects in the pitch axis) are created and separated into distinct $\lambda^\pm$ pairs akin to the dynamics of defects in two-dimensional active nematics. Initially the dynamics of this system is unsteady and resembles the turbulence of active nematics~\cite{Thampi2014b}, with $\lambda$ lines being both spontaneously created and annihilated. However, at long times the system can reach a steady state defect configuration with a fixed director field, accompanied by steady active flows. This state is discussed further in \S~\ref{sec:lambda}. 

In the case of contractile activity we do not expect the hydrodynamic modes of the cholesteric liquid crystal to be unstable from the linear calculation in \S~\ref{sec:hydroinst}. This is reflected by the simulations as the cholesteric texture is stable up to relatively high contractile activity values. An instability occurs at $\zeta = -0.01$ which is approximately two orders of magnitude larger than the threshold in the extensile case. Figure~\ref{fig:flow_dir_2D_contr} shows that above this threshold, the cholesteric order is destroyed and the resulting director field and flow field resembles that of a contractile active nematic at large activity~\cite{Sankararaman2004}. Note that if $\nu>-1$, deviations of the director along $z$ can become unstable (from the linear prediction of \S~\ref{sec:nemlimit}) and the director field can transiently display a heliconical steady state above the finite threshold. 
 
Finally, the effect of confinement is studied by considering a cell geometry in which the active cholesteric is sandwiched between two non-slip walls at the top and bottom of the simulation. Figure~\ref{fig:flow_dir_2D_ext_walls} shows the director (column (I)) and corresponding velocity field (column (II)) for three values of activity for which the system is in the active state. Interestingly, for $\zeta=0.001$ we observe a Rayleigh-Benard-like pattern of the director field with the typical undulations seen in the periodic case (see figure \ref{fig:flow_dir_2D_ext}(c)) now squeezed between two flat walls. This pattern requires a roll structure of the velocity profile, clearly visible in column (II) of figure~\ref{fig:flow_dir_2D_ext_walls}. This director pattern is accompanied by the generation of symmetric $\lambda$ line pairs, similar to the defect lattices observed in the periodic case and also to those formed in passive cholesterics undergoing a Helfrich-Hurault instability~\cite{Senyuk2006}. An increase in the activity decreases the length scale of the defect separation along the $x$-direction, sustained by shorter wavelength rolls in the flow field (figure~\ref{fig:flow_dir_2D_ext_walls}(b)). These ordered structures are destroyed for higher values of activity when the system enters the chaotic state.

\section{Lambda Lines in Active Cholesterics}
\label{sec:lambda}

Defects are a hallmark of all forms of ordered media. In two-dimensional active nematics, defects in the director field, known as disclinations, are generated spontaneously by large enough activity~\cite{Sanchez2012,Pismen2013,Giomi2013} and subsequently sustain a state of active turbulence~\cite{Thampi2014a,Thampi2014b,Giomi2015}. Disclinations of different topological charge behave differently. They are all the source of strong active flows because director distortions are necessarily large around them, but in addition disclinations of strength $+1/2$ self-propel~\cite{Narayan2007,Sanchez2012}, while those of strength $-1/2$ do not. Additionally, some experimental examples have shown that topological defects in growing bacteria and eukaryotic cell colonies in two-dimensional geometries determine the shape of the colonies as well as the spatial pattern of cell death~\cite{Kawaguchi2016,Doostmohammadi2016}. Therefore, an understanding of the motion of individual defects in active nematics and their interactions provides an explanation of many features of active nematic dynamics at high activity. Given their prominence in active nematics, it is natural to consider the behaviour of defects in active cholesterics. Extending our simulations in \S~\ref{sec:simulations}, we show that active cholesteric defects are created in an analagous way to active nematic defects, but it is also evident that their dynamics are different. However, first we consider the local structure of the active flows produced by cholesteric defects. 

\begin{figure}
\centering
\includegraphics[width=0.95\columnwidth]{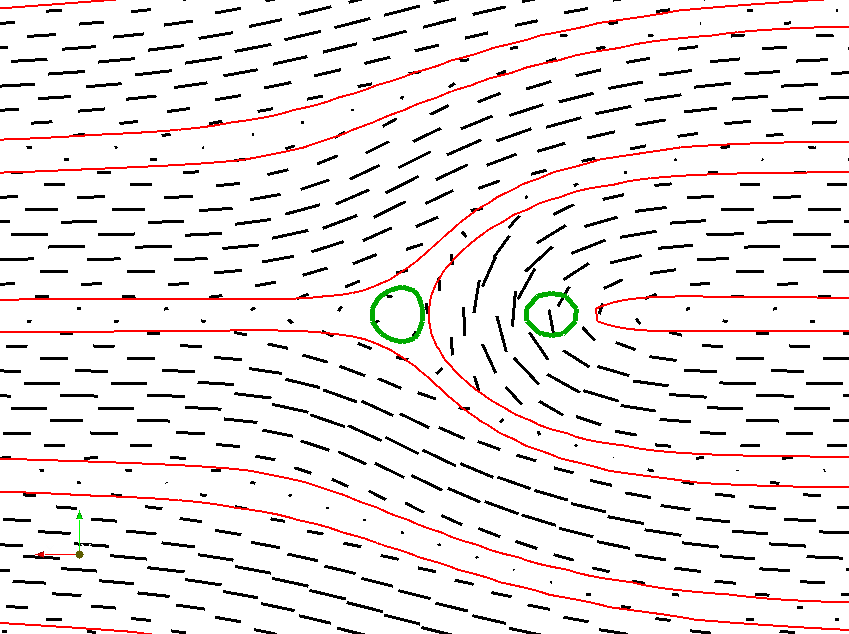}
\caption{A cholesteric dislocation, comprising a pair of $\lambda$ lines. The director field is oriented out of the page ($|n_y|>0.95$) in the regions highlighted in red. The green circles show the locations of the $\lambda^{+}$ and $\lambda^{-}$ defects.}
\label{fig:lambdaplot}
\end{figure}

The fundamental defects in cholesteric order are defects in the pitch axis known as $\lambda$ lines~\cite{deGennesProst}. In contrast to nematic disclinations, the director field is well-defined and continuous at a $\lambda$ line and instead it is the local cholesteric pitch axis that winds around the defect and is discontinuous at it. There are also disclinations in the director field in a cholesteric, known as $\chi$ lines and $\tau$ lines according to whether the pitch axis is, or is not, continuous along them, however we do not consider these here and confine our attention to $\lambda$ lines. From the cholesteric ground state, $\lambda$ lines may be produced in pairs, $\lambda^{\pm}$, with opposite winding of the pitch axis. Such a `defect-dipole' serves to create dislocations in the cholesteric layers, introducing additional full $2\pi$ rotations of the director field. An example of such, illustrating both $\lambda^{+}$ and $\lambda^{-}$ defects, is shown in figure~\ref{fig:lambdaplot}. 

\begin{figure*}
\centering
\includegraphics[width=2\columnwidth]{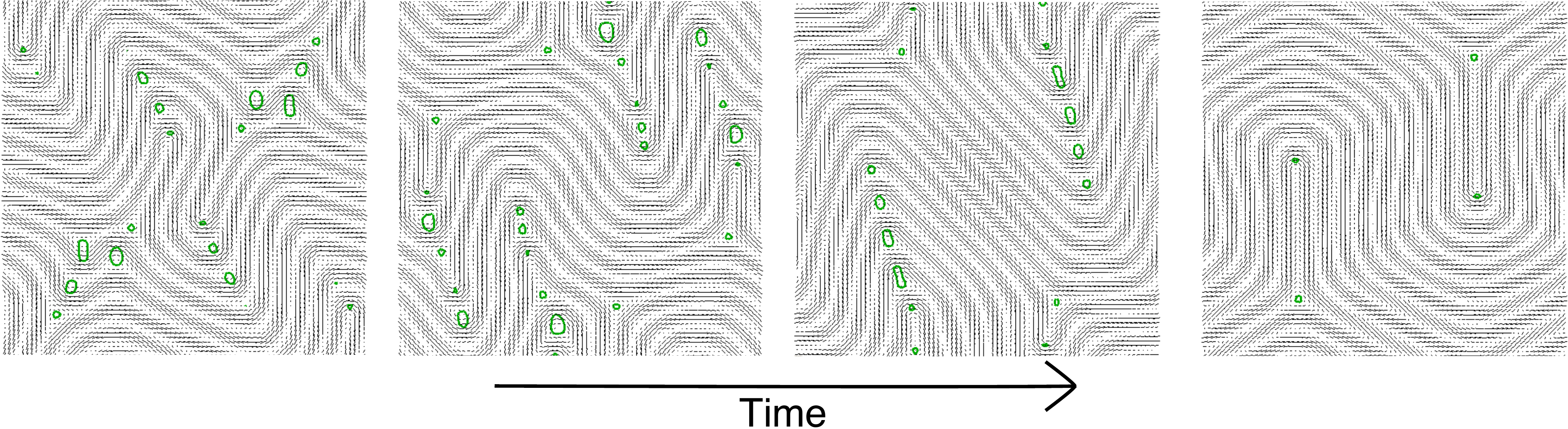}
\caption{Snapshots of the director field in an extensile active cholesteric. Lambda lines (green) are created in regions of large bend and form fixed steady state patterns. Parameters used here are the same as in previous figures except $\zeta=0.005$, $q_0 = \pi/8$, and $L_x=L_z=80$.}
\label{fig:lambdalattice}
\end{figure*}

$\lambda$ lines distort the cholesteric texture around them so that active stresses should generate flows in their vicinity, analogously to how they generate flows in the vicinity of director disclinations. To determine the local structure of these flows we need to know the structure of the director field in the vicinity of a $\lambda$ line. Following methods developed recently in~\cite{Machon2016} we show in Appendix~\ref{app:umbilic} that for a straight $\lambda$ line along the $y$-axis the director field has the local expansion 
\begin{equation}
\begin{split}
\label{dirp} \hat{\vec{n}}^{\pm} & = \biggl[ -\frac{q_0}{2}z - \alpha xz \biggr] \hat{\vec{e}}_x + \biggl[ 1-\frac{q_0^2}{8}(x^2+z^2) \biggr] \hat{\vec{e}}_y \\
& \quad + \biggl[ \frac{q_0}{2}x \pm \alpha x^2 + \frac{\alpha}{2}(x^2+z^2) \biggr] \hat{\vec{e}}_z + O(3) ,
\end{split} 
\end{equation}
where $\pm$ refers to the two types of defect ($\lambda^{\pm}$), and $O(3)$ denotes terms of cubic order or higher. The linear order terms impart the preferred local value of the twist, $\hat{\vec{n}} \cdot \nabla \times \hat{\vec{n}} = -q_0$, and no splay, $\nabla \cdot \hat{\vec{n}}=0$, along the $\lambda$ line. Its profile is encoded in the structure of the quadratic terms for which there is an arbitrary constant prefactor, $\alpha$, with dimensions of wavenumber squared. This local expansion is entirely analogous to that around umbilic points of surfaces that characterises their structure (see, {\sl e.g.}~\cite{Berry1977}) and is given here in a simplified form of the general case presented in Appendix~\ref{app:umbilic}. 

The active force density around the $\lambda$ line can be written in the form 
\begin{equation}
\begin{split}
\label{fpm} \vec{f}^{\pm} & = \frac{q_0 \zeta}{2} \biggl\{ \nabla \biggl( \frac{q_0}{4} \bigl( x^2+z^2 \bigr) + \frac{\alpha}{6} \bigl( x^3 - 3xz^2 \bigr) \biggr) \\
&  + (1\pm 1) \alpha \biggl[ \nabla \biggl( \frac{1}{3} x^3 + xz^2 \biggr) - z^2 \hat{\vec{e}}_x \biggr] \biggr\} + O(3) .
\end{split}
\end{equation}
The contributions that come from gradients can be balanced by a corresponding term in the pressure and therefore do not lead to any fluid flows. We see that the force around a $\lambda^{-}$ line is entirely of this gradient type but the same is not true for $\lambda^{+}$ lines, whose active force contains a non-gradient component directed along the $x$-axis. 
The active flows generated locally around a $\lambda^{+}$ line are therefore given approximately by 
\begin{equation}
\vec{v}^{+} \approx \frac{\alpha q_0 \zeta}{12\eta}\, z^4 \hat{\vec{e}}_x .
\end{equation} 
This local directional fluid velocity around positive strength defects, but not negative strength ones, is reminiscent of the situation for disclination lines in active nematics. The main difference is that the active force increases locally with distance from the $\lambda$ line, from being zero on it, rather than rising to a maximal value at the defect core, as is the case for nematic disclinations. This means that a $\lambda$ line in an active cholesteric does not have an intrinsic translation speed, whereas $+1/2$ active nematic defects do. 

Our numerical simulations in \S~\ref{sec:simulations} demonstrated that $\lambda$ lines are created at large extensile activities in an analogous way to $\pm1/2$ defects in nematics. The dynamics is initially unsteady, however, at long times the system can relax to a steady state configuration of $\lambda$ lines in which active and elastic stresses balance in a regular pattern of bend distortions and anti-parallel pairs of $\lambda$ lines. We show an example of this in figure~\ref{fig:lambdalattice}. This configuration is somewhat reminiscent of the defect arrangements observed in the Helfrich-Hurault instability at large applied field~\cite{Senyuk2006}, where the defects nucleate parabolic regions of circularly bent equidistant cholesteric layers that separate regions of flat layers. 

Note that fixed defect arrangements can be observed in overdamped active nematics~\cite{DeCamp2015,Doostmohammadi2016,Putzig2016}, but our analysis suggests that this can occur even in bulk cholesterics. As the activity is increased the separation between $\lambda$ line pairs is reduced, so the formation of this periodic defect lattice appears to be independent of simulation box size. At larger activity still the cholesteric dynamics are unsteady over the full simulation time, resembling the active turbulence states in nematics. In two-dimensional active nematics, defects are produced from bend `walls' in the director field, and the formation of defects transiently reinstates nematic order in the regions between them. However, the intrinsic speed of the defects means that static configurations of the director are not observed. The bend walls in nematics are analogous to the formation of strings of $\lambda$ lines seen in figure~\ref{fig:lambdalattice}. Similarly, in between the regions of large bend distortions cholesteric order is restored, but furthermore the director field can evolve to static periodic configurations. 

The nucleation of $\lambda$ lines occurs from the emergence of double twist cylinders (defects in the pitch of winding $2\pi$) in regions of large director bend, which split into $\lambda^{\pm}$ pairs oriented with the $\lambda^{+}$ flow direction acting away from the $\lambda^{-}$ defect. This is similar to the geometry of $\pm 1/2$ defects created in active nematics. However, energetic and geometric constraints inhibit the $\lambda^{\pm}$ defects from separating long distances, which is not the case for nematic defects. Defect lattices can arise in overdamped active nematics~\cite{Doostmohammadi2016}, sustained by a flow vortex lattice which matches the spatial periodicity of the defect lattice. In active cholesterics, the flow field accompanying the $\lambda$ line lattice is more complex. There is a component of the flow which is periodic over the length scale of the cholesteric layers, as well as a component of permeative flows normal to the layers which sustain the bend deformations. These permeative flows, as discussed previously, are associated with a greatly increased effective viscosity~\cite{Helfrich1969,Lubensky1972,Marenduzzo2004}, which may play a role in stabilising this defect lattice. Moreover, we have shown that the defects are not intrinsically self-propelled, which means that the flows can be balanced by elastic distortion energy of the director and so stable configurations can form in bulk.

\section{Discussion}

Active cholesterics represent a new form of active matter that combines the force-dipole stresses of actively self-propelled particles with bulk chiral ordering and structure. Their phenomenology echoes that of active nematics and smectics but is also distinctive and unique. In particular, there is a generic active instability for extensile active particles that is analogous to the contractile instability in active smectics. This leads to pitch-splay, or layer undulations, and eventually the creation of $\lambda$ lines, similar to the creation of disclinations in active nematics. However, $\lambda$ lines have different dynamics to nematic defects and do not intrinsically self-propel. The result is that the bulk active cholesteric can form stable configurations of spatially ordered $\lambda$ lines, which have only been observed previously for disclinations in overdamped active nematics. 

A number of natural directions suggest themselves for further work. For instance, our numerical studies here have only considered quasi-two-dimensional settings so that an important extension will be to perform full three-dimensional simulations to properly study the hydrodynamic pitch-splay instability and the distortion to the cholesteric ground state that it generates. In passive systems, layer undulations in the Helfrich-Hurault transition, and the analogous response to mechanical strain, produce a square lattice pattern for the distortion akin to an egg-crate structure that can evolve at higher strain into an arrangement of focal conic domains. It will be interesting to compare these with the structures that form under active stresses. 

An important feature of active nematics is the effect of confinement on the active states, in particular, the stabilisation of vortices under sufficient confinement. There are a number of different types of confinement to consider, including cell geometry and anchoring conditions, but it would also be interesting to study active cholesteric droplets, whose topology places its own constraints on the cholesteric order, leading to a wide variety of interesting textures already in passive materials. 

Our analysis of defects in active cholesterics is highly preliminary, touching only on some initial aspects of $\lambda$ lines, and there is evidently much more to be done. Defects in the director field are of considerable importance in active nematics because the active forces they generate are so large. $\chi$ lines in cholesterics should generate equally strong active forces, however, their geometry will differ from the nematic case on account of the preference for twist distortions in cholesterics, which raises an interesting question of the nature of the active flows generated by disclination lines in cholesterics.  

A major motivation for our work comes from the fact that chirality is widespread in biological materials. Many biological materials display a variety of chiral textures in addition to the simple cholesteric helix. Considering the influence of active stresses on these textures will both broaden our understanding of active materials and may help in developing concrete applications to biological systems and morphogenesis. Finally, we would like to encourage experiments aimed at developing active cholesterics; these are fascinating materials that extend and enhance our understanding of active matter.

\appendix
\section{Calculation of the dispersion relation for the hydrodynamic pitch-splay instability}
\label{app:flow}

We study the linear stability of the director perturbation given in equations \eqref{dp} and \eqref{dq}. The associated molecular field is of the form
\begin{align}
\label{hgen}  \vec{h} = \sum_n (h_n^{\perp 0}  \hat{\vec{n}}_{\perp 0} + h_n^z \hat{\vec{e}}_z) e^{i[k_x x + (k_z + nq_0)z - \omega t]}
\end{align}
where to linear order
\begin{align}
\begin{split}
\label{hp} h_n^{\perp 0} &\approx - \frac{1}{4} \Bigl\{ \bigl[ 2 ( K_1 + K_3 )k_x^2 + 4 K_2 (k_z + nq_0)^2  \bigr] \delta \phi_n  \\
& + (K_3 - K_1) k_x^2 (\delta \phi_{n+2} + \delta \phi_{n-2})    \vphantom{\Bigl(}\\
& - 2(K_1+K_3)ik_xq_0 ( \delta \theta_{n+1} +  \delta \theta_{n-1} )    \vphantom{\Bigl(}\\
& + 2 (K_2 - K_1) i k_x (k_z + nq_0) ( \delta \theta_{n+1} -  \delta \theta_{n-1} )\Bigr\} ,
\end{split}
\end{align}
\begin{align}
\begin{split}
\label{hz} h_n^z &\approx - \frac{1}{4} \Bigl\{ \bigl[ 2 ( K_2 + K_3 )k_x^2 + 4 K_1 (k_z + nq_0)^2 + 4K_3 q_0^2 \bigr] \delta \theta_n\\
& + (K_3 - K_2) k_x^2 (\delta \theta_{n+2} + \delta \theta_{n-2})     \vphantom{\Bigl()}\\
& + 2(K_2 + K_3) ik_xq_0 (\delta \phi_{n+1} + \delta \phi_{n-1})      \vphantom{\Bigl()}\\
& + 2(K_2 - K_1) ik_x(k_z + nq_0) (\delta \phi_{n+1} - \delta \phi_{n-1}) \Bigr \} \, .
\end{split}
\end{align}
Also to linear order the distortion force in the Stokes equations can be given in terms of the molecular field as
\begin{align}
\begin{split}
\label{fd} \vec{f}^d_n \approx& \frac{1}{4}\biggl\{ \Bigl[ ik_x(\nu-1) (h_{n+1}^z + h_{n-1}^z) - 4q_0 h_n^{\perp 0} \Bigr]\hat{\vec{e}}_z \\
& + \Bigl[ -2ik_x h_n^{\perp 0} + \nu ik_x(h_{n+2}^{\perp 0} + h_{n-2}^{\perp 0})\\
& \;\quad - (\nu+1)(k_z+nq_0)(h_{n+1}^z-h_{n-1}^z) \Bigr]\hat{\vec{e}}_y\\
& + \Bigl[\nu k_x (h_{n+2}^{\perp 0} - h_{n-2}^{\perp 0}) \\
& \;\quad + i (\nu+1) (k_z + n q_0)(h_{n+1}^z + h_{n-1}^z) \Bigr] \hat{\vec{e}}_x \biggr\} \, .
\end{split}
\end{align}

The Stokes equations are solved for the flow by separating into components parallel to $\vec{e}_y$ and to $\vec{k}_\perp = (k_z + nq_0)\hat{\vec{e}}_x - k_x \hat{\vec{e}}_z$ (thus projecting out the pressure) with the third equation given by incompressibility $\nabla \cdot \vec{v}=0$. The velocity solution takes the form:
\begin{align}
\begin{split}
\label{vn} & \vec{v}_n = \frac{(k_z+nq_0)\bigl[ (k_z + nq_0)f_n^{(x)} - k_xf_n^{(z)} \bigr]}{\eta\bigl[k_x^2+(k_z+nq_0)^2\bigr]^2}\hat{\vec{e}}_x \\
& + \frac{f_n^{(y)}\hat{\vec{e}}_y}{\eta\bigl[k_x^2+(k_z+nq_0)^2\bigr]} + \frac{k_x\bigl[ k_xf_n^{(z)} - (k_z + nq_0)f_n^{(x)} \bigr]}{\eta\bigl[k_x^2+(k_z+nq_0)^2\bigr]^2}\hat{\vec{e}}_z
\end{split}
\end{align}
where $\vec{f}_n = \vec{f}_n^d + \vec{f}_n^a$ and the superscripts $(x,y,z)$ denote Cartesian components of $\vec{f}$. 

Finally, the director dynamic equations \eqref{dndt} give the stability conditions for the set of perturbations $\delta \phi_n$ and $\delta \theta_n$, equations \eqref{dpdt} and \eqref{dqdt}. In general each mode $n$ is coupled to modes $n\pm6$ through the passive flow terms. We take the hydrodynamic limit $k \ll q_0$ and expand the relations \eqref{dpdt} and \eqref{dqdt} as series expansions. For $|n| > 2$ the frequency $\omega$ is dominated by diagonal terms which strongly damp the perturbations. Thus these modes can be incorporated as small perturbations to the equations for $\delta \phi_0$, $\delta \theta_{\pm1}$ and $\delta \phi_{\pm 2}$ by repeated substitution of equations \eqref{dpdt} and \eqref{dqdt} for  these modes. This reduces the calculation to a $5\times5$ eigenvalue problem of the form $(\vec{A} + i\omega \mathbb{I}) \vec{x} = 0$ where $\vec{x} = (\delta \phi_0, \delta \theta_1, \delta \theta_{-1}, \delta \phi_2, \delta \phi_{-2})$ and the structure of the matrix $\vec{A}$ is 
\begin{equation}
\vec{A} = 
\begin{pmatrix}
O(k^0) & O(k^{-1}) & O(k^{-1}) & O(k^{-1}) & O(k^{-1})\\
O(k^1) & O(k^{0}) & O(k^{0}) & O(k^{0}) & O(k^{0})\\
O(k^1) & O(k^{0}) & O(k^{0}) & O(k^{0}) & O(k^{0})\\
O(k^1) & O(k^{0}) & O(k^{0}) & O(k^{0}) & O(k^{0})\\
O(k^1) & O(k^{0}) & O(k^{0}) & O(k^{0}) & O(k^{0})
\end{pmatrix} . 
\end{equation}
Clearly these five modes are, in general, coupled at lowest order in $k$. The resulting characteristic equation for $\omega$ suggests that all solutions $\omega$ are of order $k^0$ (even when $\zeta = 0$). However, when $k_x = 0$ the matrix is completely diagonalised and the only diffusive mode corresponds to $\delta \phi_0$ and this is unaffected by the activity. in this limit the $\delta \theta$ modes have a finite stability threshold for the activity, and is of the form calculated in section \ref{sec:nemlimit}. 

If $k_z = 0$ then the modes $\delta \phi_{\pm 2}$ decouple from the other three at order $k^0$ and so the problem can be reduced to a $3\times3$ problem again by accounting for the contributions of all other modes by repeated substitution of the relevant modes of equations \eqref{dpdt} and \eqref{dqdt}. This reduced eigenvalue problem then can be written to order $k^2$ as the solution to $(\vec{A}' + i\omega) \vec{x}' = 0$ where $\vec{x}' = (\delta \phi_0, \delta \theta_1, \delta \theta_{-1})$ and 
\begin{equation}
\label{eq:Aprime} \vec{A}' =
 \begin{pmatrix}
 a_0 + a_2 k^2  & i k^{-1} (b_{-1} + b_1 k^2) & i k^{-1} (b_{-1} + b_1 k^2) \\
 i k (c_1 + c_3 k^2) & d_0 + d_2 k^2 & e_0 + e_2 k^2 \\
 i k (c_1 + c_3 k^2) & e_0 + e_2 k^2 & d_0 + d_2 k^2
 \end{pmatrix} .
\end{equation}
The constants $a_0$, \ldots, $e_2$ are cumbersome in their length; we give them in the one elastic constant approximation at the end of this appendix. To order $k^0$ the eigenfrequencies are 
\begin{align}
\begin{split}
\omega_0 & = \frac{i}{4} \biggl\{ - \biggl( \frac{\zeta\nu}{\eta} +  \frac{\tilde{K}q_0^2}{\tilde{\eta}} \biggr) + \biggl[ \biggl( \frac{\zeta\nu}{\eta} + \frac{\tilde{K}q_0^2}{\tilde{\eta}} \biggr)^2\\
& \qquad - \frac{2\zeta}{\eta}\biggl( \frac{\zeta(1+\nu)}{\eta} + \frac{\tilde{K}q_0^2}{\eta'} \biggr) \biggr]^{1/2} \biggr\} + O(k^2) ,
\end{split} \\
\begin{split}
\omega_1 & = - \frac{i}{4} \biggl\{ \biggl( \frac{\zeta\nu}{\eta} + \frac{\tilde{K}q_0^2}{\tilde{\eta}} \biggr) + \biggl[ \biggl( \frac{\zeta\nu}{\eta} + \frac{\tilde{K}q_0^2}{\tilde{\eta}} \biggr)^2\\
& \qquad - \frac{2\zeta}{\eta}\biggl( \frac{\zeta(1+\nu)}{\eta} + \frac{\tilde{K}q_0^2}{\eta'} \biggr) \biggr]^{1/2} \biggr\} + O(k^2) ,
\end{split} \\
\omega_2 & = -\frac{i}{4} \biggl[ \frac{\zeta(1+\nu)}{\eta} + \frac{\tilde{K}q_0^2}{\eta'} \biggr] + O(k^2) ,
\end{align}
where $\tilde{K} = (K_1+K_3)/2$, $\tilde{\eta} = (4/\gamma + (1+\nu)^2/\eta)^{-1}$ and $\eta' = (8/\gamma + (1+\nu)^2/\eta)^{-1}$. In the passive limit ($\zeta = 0$) $\omega_0 = O(k^2)$ signifying that this is the hydrodynamic mode. To calculate the leading order contribution in the passive limit, we take $\zeta = 0$ and look for solutions to the characteristic equation at order $k^2$ assuming $\omega = D_0 k^2$. In this case the characteristic equation becomes:
\begin{align}
\begin{split}
D_0 & = \frac{i}{q_0 \tilde{\eta}} \biggl[ (b_1 + a_2 q_0) \biggl( \frac{4}{\gamma} + \nu \frac{1+\nu}{\eta} \biggr) \\
& \qquad + \bigl( d_2 + e_2 - 2c_3q_0^2 \bigr) \frac{1+\nu}{\eta} \biggr] + O(k^2) ,
\end{split} \\
\label{diff0} & \approx - \frac{i 3 K_3}{16 \eta} \frac{\tilde{\eta}}{\eta'} \, .
\end{align}
Expanding $\omega_0$ to first order in the activity and adding the passive diffusive contribution then gives the expression for $\omega_0$ given in equation \eqref{omega0} in the main text. In the long wavelength limit the eigenfunction corresponding to this mode is
\begin{align}
\delta \theta_{\pm 1} = - \frac{i k}{2 q_0}\biggl[ \frac{\tilde{K}q_0^2\eta + \zeta(1+\nu)\bar{\eta}/2}{\tilde{K}q_0^2\eta + \zeta\nu\bar{\eta}} \biggr]\delta \phi_0
\end{align}
where $\bar{\eta} = (4/\gamma + \nu(1+\nu)/\eta)^{-1}$. This eigenmode reduces to the pitch splay perturbation of equations \eqref{eigf1} and \eqref{eigf2} in the limit of small activity.

Working to order $k^2$ for the hydrodynamic mode, the coefficients $a_0, \dots, e_2$ appearing in the matrix $\vec{A}^{\prime}$, equation~\eqref{eq:Aprime}, are given in a one elastic constant approximation as follows:  
\begin{equation}
a_0 = - \frac{K q_0^2 (1+\nu)}{2\eta} , 
\end{equation}
\begin{equation}
\begin{split}
a_2 & = - \frac{\zeta  \nu}{8 \eta q_0^2} - K\Bigl[ \frac{1}{8 \eta} (\nu ^2 + \nu + 2) + \frac{1}{\gamma} \Bigr] \\
& \quad - \frac{ \gamma (\zeta +K (\nu +1) q_0^2)(\zeta +K (2 \nu +1) q_0^2)}{4 \eta q_0^2 (\gamma \zeta \nu + 2 K q_0^2 (\gamma \nu^2 + 8\eta))} ,
\end{split} 
\end{equation}
\begin{equation}
b_{-1} = \frac{q_0}{2\eta} \bigl[ \zeta + K q_0^2 (1+\nu) \bigr] , 
\end{equation}
\begin{equation}
\begin{split}
b_1 & = i \Bigl\{ \zeta^2 G (\nu^2 - 2) + K^2 q_0^4 \bigl[ 128 G^{-1} + 8 (4 \nu^2 + 5 \nu +1) \\
& + G (2 \nu^4 + 5 \nu^3 - 3 \nu^2 - 6 \nu -2) \bigr] \\
& + \zeta K q_0^2 \bigl[ G (3\nu^3 + 3\nu^2 - 6\nu - 4) + 8(3 \nu + 1) \bigr] \Bigr\} \Big/ \\
& \Bigl\{ 8 q_0 \bigl[ \zeta \gamma \nu + 2 K q_0^2 (\gamma \nu^2 + 8 \eta) \bigr] \Bigr\} ,
\end{split} 
\end{equation}
\begin{equation}
c_1 = - \frac{\zeta (1+\nu)}{8 q_0 \eta} - \frac{K q_0}{4\bar{\eta}} , 
\end{equation}
\begin{equation}
\begin{split}
c_3 & = i \Bigl\{ 2 \zeta^2 G \nu (2\nu + 1) + 2 K^2 q_0^4 \bigl[ 8 (\nu^2 - 2\nu - 1) \\
& + G (\nu^4 - 6 \nu^3 + 13 \nu^2 + 16 \nu + 4) \bigr] \\
& + \zeta K q_0^2 \bigl[ G (9 \nu^3 - 2 \nu^2 + 17 \nu + 8) + 64 \nu) \bigr] \Bigr\} \Big/ \\
& \Bigl\{ 64 q_0^3 \bigl[ \zeta \gamma \nu + 2 K q_0^2 (\gamma \nu^2 + 8 \eta) \bigr] \Bigr\} ,
\end{split} 
\end{equation}
\begin{equation}
d_0 = -\frac{\zeta (3\nu + 1)}{8\eta} - 2 K q_0^2 \biggl[ \frac{1}{\gamma} + \frac{(3\nu+1)(\nu+1)}{\eta} \biggr] , 
\end{equation}
\begin{equation}
\begin{split}
d_2 & = - \Bigl\{ 2 K^2 q_0^4 \bigl[ 1536 G^{-1} + 32 (9 \nu^2 + 4 \nu + 11) \\
& + 4 G (3 \nu^4 + 10 \nu^3 - 5 \nu^2 - 26 \nu -10) \\
& + G^2 \nu^2 (\nu^3 - \nu^2 - 5 \nu - 3) \bigr] \\
& + \zeta K q_0^2 \bigl[ G^2 \nu (3 \nu^3 - 5 \nu^2 - 11 \nu - 3) \\
& - 4 G (15 \nu^3 + 17 \nu + 20) - 32 (13 \nu + 4) \bigr] \\
& + \zeta^2 G \nu \bigl[ G (\nu - 3) (\nu + 1) - 8 (5 \nu + 2) \bigr] \Bigr\} \Big/ \\
& \Bigl\{ 256 q_0^2 \bigl[ \zeta \gamma \nu + 2 K q_0^2 (\gamma \nu^2 + 8 \eta) \bigr] \Bigr\} ,
\end{split} 
\end{equation}
\begin{equation}
e_0 = \frac{1-\nu}{8\eta} \bigl[ \zeta + K q_0^2 (1+\nu) \bigr] , 
\end{equation}
\begin{equation}
\begin{split}
e_2 & = \Bigl\{ \zeta^2 G \nu \bigl[ G (\nu - 3) (\nu + 1) + 8 \bigr] \\
& + 2 K^2 q_0^4 \bigl[ G^2 \nu^2 (\nu - 3) (\nu + 1)^2 \\
& - 4 G (7 \nu^4 + 2 \nu^3 - 9 \nu^2 - 6 \nu + 2) - 32 (7 \nu^2 + 3) \bigr] \\
& - \zeta K q_0^2 \bigl[ 32 \nu + 4 G (9 \nu^3 - 8 \nu^2 - 9 \nu + 4) \\
& - G^2 \nu (3 \nu^3 - 5 \nu^2 - 11 \nu - 3) \bigr] \Bigr\} \Big/ \\
& \Bigl\{ 256 q_0^2 \bigl[ \zeta \gamma \nu + 2 K q_0^2 (\gamma \nu^2 + 8 \eta) \bigr] \Bigr\} ,
\end{split}
\end{equation}
where $G=\gamma\eta^{-1}$.

\section{Director field in the vicinity of a $\lambda$ line}
\label{app:umbilic}

The $\lambda^{\pm}$ defects in a cholesteric are examples of generic umbilic lines~\cite{Machon2016}, which are analogous in structure to the umbilic points of degenerate principal curvature on surfaces (see, {\sl e.g.}~\cite{Berry1977}). If the director field is the normal to a family of surfaces, as in a smectic, then the definitions are equivalent. In the general case umbilic lines are identified by degeneracies in the derivatives of the director field along directions that are orthogonal to it. Concretely, in a local orthonormal frame $\{\hat{\vec{d}}_1,\hat{\vec{d}}_2,\hat{\vec{n}}\}$, $\lambda$ lines may be identified by the vanishing of the matrix~\cite{Machon2016} 
\begin{equation}
\label{Delta} \Delta = \begin{pmatrix} \Delta_1 & \Delta_2 \\ \Delta_2 & - \Delta_1 \end{pmatrix} ,
\end{equation}
whose components are 
\begin{align}
\Delta_1 & = \frac{1}{2} \Bigl[ \hat{\vec{d}}_1 \cdot (\nabla \hat{\vec{n}}) \cdot \hat{\vec{d}}_1 - \hat{\vec{d}}_2 \cdot (\nabla \hat{\vec{n}}) \cdot \hat{\vec{d}}_2 \Bigr] , \\
\Delta_2 & = \frac{1}{2} \Bigl[ \hat{\vec{d}}_1 \cdot (\nabla \hat{\vec{n}}) \cdot \hat{\vec{d}}_2 + \hat{\vec{d}}_2 \cdot (\nabla \hat{\vec{n}}) \cdot \hat{\vec{d}}_1 \Bigr] .
\end{align}
This definition was applied to identify $\lambda$ lines in the simulations shown in \S~\ref{sec:lambda}. 

To give an analytical expression for the active flows generated around a $\lambda$ line we require an expression for the director field in the vicinity of the defect. We begin with a generic expansion of the director field about the origin, to quadratic order, assuming that $n_y=1$ at $(x,z)=(0,0)$ and that there is no dependence on the direction $y$ (as in our quasi-two-dimensional simulations): 
\begin{equation}
\begin{split}
\hat{\vec{n}} & = \Bigl[ k_0 x + k_1 z + \alpha_0 x^2 + \alpha_1 xz + \alpha_2 z^2 \Bigr] \hat{\vec{e}}_x \\
& \quad + \biggl[ 1 - \frac{1}{2} \bigl( k_0 x + k_1 z \bigr)^2 - \frac{1}{2} \bigl( m_0 x + m_1 z \bigr)^2 \biggr] \hat{\vec{e}}_y \\
& \quad + \Bigl[ m_0 x + m_1 z + \beta_0 x^2 + \beta_1 xz + \beta_2 z^2 \Bigr] \hat{\vec{e}}_z . \\
\end{split}
\end{equation} 
Firstly, for the origin to be umbilic, the matrix~\eqref{Delta} must vanish there, giving the conditions $m_0=-k_1$ and $m_1=k_0$. Next, we impose that the director should be divergence free at equilibrium to minimise the splay term in the free energy, which gives $k_0=0$, $\alpha_1=-2\beta_2$ and $\beta_1=-2\alpha_0$. These conditions give the generic local form of the director near a straight $\lambda$ line in a cholesteric. 

The matrix \eqref{Delta} winds by $\pm 2\pi$ about its degeneracy at the origin. Denoting by $\theta$ an angular coordinate on a small loop encircling the $\lambda$ line, we can write $\Delta_1 = \alpha \cos(\theta+\phi)$ and $\Delta_2 = \pm \alpha \sin(\theta+\phi)$, where $\alpha$ sets the magnitude of $\Delta$ and $\phi$ sets the in-plane orientation of the defect. The $\pm$ sign corresponds to a positive or negative winding of $\Delta$ and hence to the sign of the $\lambda^{\pm}$ defect. For a $\lambda^{+}$ line this gives the conditions $\alpha_2=3\alpha_0=3\alpha\cos\phi/2$ and $\beta_0=3\beta_2=3\alpha\sin\phi/2$, and for a $\lambda^{-}$ line it gives $\alpha_2=-\alpha_0=\alpha\cos\phi/2$ and $\beta_0=-\beta_2=-\alpha\sin\phi/2$. The orientation $\phi$ is arbitrary and here we set it to $\phi=\pi/2$. 

Finally, the twist of this director field at the origin ({\sl i.e.} along the $\lambda$ line) is $\hat{\vec{n}}\cdot\nabla\times \hat{\vec{n}} = 2k_1$. We set $k_1=-q_0/2$ to correspond to the preferred local value. This results in the director field \eqref{dirp}.

\begin{acknowledgement}
We acknowledge funding from the UK EPSRC through Grant Nos. EP/N007883/1 (CAW and GPA) and EP/J007404/1 (AT and DM). SR was supported in part by a J C Bose Fellowship of the SERB, India. 
\end{acknowledgement}

\end{document}